\documentclass[aps,prl,preprint,superscriptaddress]{revtex4}

\usepackage[dvips]{graphicx}
\usepackage[]{amsmath}
\usepackage[]{latexsym}

\newcommand{\mb}[1]{\mbox{\boldmath $#1$}}

\newcommand{\eas}[0]{\begin{eqnarray*}}
\newcommand{\eae}[0]{\end{eqnarray*}}
\newcommand{\les}[0]{\begin{equation}}
\newcommand{\lee}[0]{\end{equation}}
\newcommand{\leas}[0]{\begin{eqnarray}}
\newcommand{\leae}[0]{\end{eqnarray}}
\newcommand{\mat}[4]
{
\left(
\begin{array}{cc}
#1 & #2 \\
#3 & #4 
\end{array}
\right)
}
\newcommand{\mvec}[2]
{
\left(
\begin{array}{c}
#1  \\
#2  
\end{array}
\right)
}
\begin{document}


\title{Topological Quantum Phase Transitions 
in Superconductivity
 on Lattices 
} 


\author{Y. Hatsugai}
\email[]{hatsugai@pothos.t.u-tokyo.ac.jp}
\affiliation{Dept. of Applied Physics, Univ. of Tokyo}
\affiliation{PRESTO JST}
\author{S. Ryu}
\affiliation{Dept. of Applied Physics, Univ. of Tokyo}


\date{Nov.1, 2001}

\begin{abstract}

Topological quantum phase transitions in superconductivity are
discussed on two dimensional lattices. 
The main focus is on the Chern number for superconducting states. 
Each superconductivity is characterized by the Chern number,
and
the quantum phase  transition is associated with topological changes
of the quasiparticle Bloch function in the Brillouin zone.
For the superconducting case, the Chern number
has several equivalent but different topological expressions
given by vortices, the Dirac monopole and strings.
 We demonstrate quantum phase transitions by these 
topological quantities both for singlet and triplet cases.
\end{abstract}


\maketitle


The quantum phase transition is a drastic change of a ground state
when physical parameters of a system are varied.
This is a  phase transition at zero temperature
with quantum fluctuation
which raises many interesting physical questions.
Recently substantial number of works have focused on the topic reflecting
its importance. 
Typical examples
include the Mott transition in strongly correlated systems and the 
Hall plateau transition in the quantized Hall effect
 \cite{kl-qhe,float-laughlin,float-bhat,yh-random-ll}. 
The latter is special in the sense that the ground state is
characterized by a quantized quantity 
( quantized Hall conductance $\sigma_{xy}$ ).
As is known today, 
its quantization originates from a topological character of the Hall
conductance\cite{edge_l,tknn,mkann,chern_yh}. 
Then the transition is governed by topological objects such as 
 the Chern numbers, vortices and  edges states\cite{tknn,mkann,edge_yh,chern_yh}. 
Each phase is characterized by the 
different topological quantum number 
although the symmetry of the state is not necessarily  different.
It is quite different from usual phase transitions
and is characteristic to the topological quantum 
phase transition\cite{wen_topological1}.
There have been several features for this type of the phase
transition. 
Some of them are a special form of a selection rule, and 
the possible stability of the phase
which puts special importance on the topological transitions 
and discriminates from other 
quantum phase transitions\cite{wen_topological1,yh-random-ll}. 

Recently, there have been trials to extend the concept
of this topological phase transition to unconventional singlet 
superconductivity\cite{kaga,ym-yh-pairing,read-green,smf}. 
In the discussions, mapping the
system into the standard quantum Hall
system is essential. 
Here the ``spin'' Hall conductance plays a main role
which is given by the Chern numbers. 
A non trivial example of such states is given by time-reversal
symmetry breaking superconductivity
\cite{volovik1,laughlin-dd,ym-yh-pairing,read-green,smf}. 
Leaving its reality apart, the topological phase transition in 
superconductivity raises important theoretical questions 
how the topological character restricts transition types. 
In this paper, we take the generalized Bogoliuvov-de Gennes (B-dG)
hamiltonian for superconductivity to discuss the
topological quantum phase transition and demonstrate the topological
character using several topological expressions.

Search for spin triplet superconductivity has also long history
but several important materials have been found recently which raised 
intense re-investigations of triplet superconductivity.
One of the interesting aspects  of
the triplet state is its rich structure of the order parameters
which allows the existence of non-trivial topological structures. 
On this point, the topological characters of the 
chiral $p$-wave superconductivity is
investigated by several groups\cite{smf,read-green,goryo-p,furusaki-p}. 
 In this paper, 
we treat this spin triplet
case on a two dimensional lattice based on the
same strategy as the spin singlet case.
Then we can characterize the triplet state  by the Chern
number and several topological expressions as well \cite{sr-yh-sup}.

Comparing to the usual Quantum Hall effect with many Landau bands,
there are only 
two quasiparticle bands of quasiparticles ( and quasiholes )
in the superconductivity. 
This special but simple situation  allows us to use 
the interesting Berry's parameterization.
Then the gauge fixing to calculate the explicit Chern number
also produces an interesting topological expression.
Using this gauge fixing, we establish a relation between 
several different topological expressions.

{\it{The Chern number}\ }
 To demonstrate the topological transition in superconductivity,
we investigate systems described by the following 
B-dG hamiltonian for superconducting quasiparticles
\begin{eqnarray*} 
H
 &=& 
 \sum_{ij} (
t_{ij}^\sigma 
 c_{i\sigma}^\dagger c_{j\sigma}
 + 
\Delta_{ij} 
c_{i\uparrow}^\dagger c_{j\downarrow}^\dagger
+ 
\Delta_{ij} ^*
c_{j\downarrow} c_{i\uparrow})
-\mu \sum_{i\sigma} c_{i\sigma}^\dagger c_{i\sigma}.
\end{eqnarray*} 
In this letter, 
we assume the system is translationally invariant
as $t^\sigma_{ij}=t^\sigma(i-j)$ and $\Delta_{ij}=\Delta(i-j)$.
Then the hamiltonian is
written as 
\[
H =  \sum_k 
\mb  {c}^\dagger (\mb{k} ) \mb{h} (\mb{k} )\mb  {c}, \ \ \
\mb{h} (\mb{k} ) = 
\mat
{\epsilon^\uparrow( \mb{k}) }
{\Delta(\mb{ k})}
{\Delta^*(\mb{k} )}
{-\epsilon^\downarrow(- \mb{k}) } 
\]
where 
$
\mb  {c}^\dagger (\mb{k} )
=
(c^\dagger _\uparrow(\mb{k}),c_\downarrow(\mb{k}) )
$,
$\epsilon^\sigma( \mb{k}) 
= 
\sum_j  e^{-i \mb{k} \cdot \mb{r}_j} t^\sigma(j)$ and
$\Delta( \mb{k}) 
= 
\sum_j  e^{-i \mb{k} \cdot \mb{r}_j} \Delta(j)
$.
We further require $t_{ij}^\uparrow=t_{ji}^\downarrow$.
Then we have $\epsilon^\uparrow(\mb{k})
=\epsilon^\downarrow(-\mb{k})\equiv\epsilon(\mb{k})$.
Generically various realizations of order parameters
are determined by the self-consistent equation.
Here let us assume a form of the order parameters to discuss 
topological quantum phase transitions. 

The ``spin'' Hall conductance of the 
superconducting state on a lattice is given by
 the generalized TKNN formula 
as
$
\sigma = \frac {e^2}{h} C
$ 
where $ C  \equiv  \frac {1}{2\pi i} \int_{T^2} d \mb{S}_k\,\cdot
{\rm rot } _k \,   \mb{A} _k
$,
$
\mb{A}_k=  \langle \mb{k}  | \mb{\nabla }_k
 \mb{k} \rangle 
$ \cite{tknn,mkann,chern_yh,kaga,ym-yh-pairing,read-green}.
The Chern number $C $, is given by a total vorticity of the 
quasiparticle Bloch function
$|\mb{k} \rangle $ for the {\it negative energy} quasiparticle,
$ \mb{h}(\mb{k} ) |  \mb{k} \rangle  =  -E(\mb{k} )|\mb{k} \rangle $,
$E(\mb{k} )=\sqrt{\epsilon(\mb{k} )^2+| \Delta (\mb{k} )|^2  }$
since the quantum mechanical average is taken 
 over the grand canonical ensemble. 
The integration is over the Brillouin zone which is topologically a
torus $T^2$, and $ d\mb{S}=(dk_x,0,0)\times (0,dk_y,0)=(0,0,dk_xdk_y)$ is
an infinitesimal area.
The Chern number
has an apparent topological expression which is written as a winding number
of a relative phase of the Bloch function\cite{mkann,chern_yh}
\[
C = -N_{\rm winding} \equiv
-\sum_\ell \frac {1}{2\pi}\oint_{\partial R_\ell} d \mb{k} \cdot 
\mb{\nabla} {\rm Im\ }  \log 
\frac 
{ |\mb{k}\rangle_1}
{ |\mb{k}\rangle_2}.
\]
The summation is over the poles of 
$\frac 
{ |\mb{k}\rangle_1}
{ |\mb{k}\rangle_2}
$
(zeros of the the second component
of the Bloch function ${ |\mb{k}\rangle_2}$) in 
the Brillouin zone ($R_\ell$ is a small area around each pole)
\cite{mkann,chern_yh}. 

By the Berry's parametrization\cite{berry}, 
the hamiltonian $\mb{h}(\mb{k})$ can be written as
\[
\mb{h}(\mb{k} ) 
= \mb{R} (\mb{k})  \cdot \mb{\sigma}  
\]
where 
$\mb{\sigma}=(\sigma_x,\sigma_y,\sigma_z)$  are  the Pauli
matrices 
and 
$ \mb{R} = (R_x,R_y,R_z)=
(
{\rm Re\, }\Delta(\mb{k}),
-{\rm Im\, }\Delta(\mb{k}),
\epsilon(\mb{k}) )
$.
Then the expression for the Chern number is rewritten by these coordinates
in $\mb{R}$ space as
\[
C = 
\frac {1}{2\pi i} \int_{R(T^2)} d \mb{S}_R\,\cdot
{\rm rot } _R \,  \mb{A} _R
\]
where
$\mb{A}_R =    \langle \mb{R} | \mb{\nabla }_R
\mb{R}\rangle 
$
and 
$
|\mb{k}\rangle = |\mb{R}(\mb{k}) \rangle 
$.
In a particular gauge ( a phase convention of the vector),
 it is written as
$
| \mb{R}\rangle  = 
\mvec
{\sin \frac { \theta }{2} }
{e^{i\phi}\cos \frac { \theta }{2} }
$
where $(R,\theta,\phi)$ is a polar coordinate of $\mb{R} $.
The integral region $R(T^2)$ is 
a closed surface in three dimensional parameter space
mapped from the Brillouin zone $T^2$.
As is well known, the 
corresponding vector potential defines a magnetic monopole
at the origin in the three dimensional space as
$
{\rm div}\,  {\rm rot}\,  \mb{A} _R =  - {2\pi i} \, \delta_R(\mb{R} )
$ \cite{dirac,wu-yang,berry}.
Therefore, by the Gauss' theorem, 
we have another expression for the Chern number as
\[
C =  \int_{R(T^2 )} dV  \delta_R(\mb{R} )= -N(R(T^2),O)= -N_{covering}
\]
where $N_{covering}=N(R(T^2),O) $ is a degree of covering by the closed
surface
$R(T^2)$ around the origin $O$. 
This is another topological expression characterizing the negative
energy band.

Now let us calculate the winding number $N_{winding}$ by this
gauge. In the present gauge, 
$
\mb{\nabla}_k 
{\rm Im\ }  \log 
\frac
{ |\mb{k}\rangle_1}
{ |\mb{k}\rangle_2}
=
-\mb{ \nabla}_k \phi
$ and  we have
$
N_{winding} 
= 
\frac {1}{2\pi} \sum_{P_-}
\oint_{R(\partial R_\ell)} d \mb{R} \cdot \mb{\nabla} _R \phi.
$
Since the zeros of the second component of the 
Bloch function are defined by $ \cos\theta/2=0 $,
we rewrite the above equation as
\begin{eqnarray*} 
C &=& -N_{winding}=-  \sum_{P_-} I({R(T^2)},z_-)
\end{eqnarray*} 
where $z_-$ is a negative $z$-axes in the $\mb{R}$ space
(the Dirac string)
and $P_-$'s are intersection points of the closed surface
${R(T^2)}$ and the Dirac string $z_-$.
The integer $I ({R(T^2)},z_-)$ 
is an intersection number
which gives the multiplicity of the local covering
at $P_-$'s.
It is a positive integer
when  the Dirac string $z_-$ is parallel to the normal 
vector of the surface 
$
\frac {\partial \mb{R}  }{\partial k_x } 
\times
\frac {\partial \mb{R}  }{\partial k_y } 
$
and a negative integer when it is anti-parallel.
$N_{winding}$
 is apparently integral and zero if the origin is 
outside the surface. 
It is one of the advantages of the present expression. 

Now we have established a relation for the two topological 
integers as 
\begin{eqnarray*} 
 N_{covering} &=& N_{winding} \\
N(R(T^2),O) &=&  \sum_{P_-} I({R(T^2)},z_-).
\end{eqnarray*} 
This relation is physically apparent but it confirms the 
consistency of the discussion. 

{\it{Demonstration}\ }
 As is discussed above, 
the  Chern number is a topological quantity and is stable against
 small perturbation. 
Each superconducting state is characterized by
an integer and the topological quantum transition is associated
 with topological change of geometrical objects,
one of which 
is the relative phases of the Bloch functions of the quasiparticles
and 
the others are the closed surface 
$R(T^2)$ and the Dirac string $z_-$. 
The topological transition of the superconductivity is 
clearly observed by investigating these topological objects.

Now let us take following two examples of the order parameters
to demonstrate the topological transition. 
\begin{eqnarray*} 
{\rm Case\ I}: && {\rm singlet }, \Delta_{ij}=\Delta_{ji},\ 
  \\
&&\Delta_{i,i} =\Delta_0,\ \Delta_{i,i+\hat x} =
 -\Delta_{i,i+\hat y} 
= \Delta_{x^2-y^2} \\
&&\Delta_{i,i+\hat x+\hat y} =
-\Delta_{i+\hat x,i+\hat y} = i \Delta_{xy} \\
{\rm Case\ II}: && {\rm triplet }, \Delta_{ij}=-\Delta_{ji},\ 
 \\
&& \Delta_{i,i+\hat x} 
=
-i \Delta_{i,i+\hat y} 
=
 \Delta_{x} 
\end{eqnarray*} 
where $\Delta_{x^2-y^2}$ and $\Delta_{xy}$ are real. 
The hopping $t^\sigma_{ij}=t$, is also 
real and only nonzero on the nearest neighbor links.
We have
$\epsilon(\mb{k})=-2 t (\cos k_x+\cos k_y)-\mu$,
for the both cases,
and
 $
\Delta (\mb{k} ) = \Delta_0+
2 \Delta_{x^2-y^2} ( \cos k_x-\cos k_y)
+
2i \Delta_{xy} (\cos (k_x+k_y)-\cos(k_x- k_y))
$ for the singlet case (Case I)
and 
 $
\Delta (\mb{k} ) = 
2i \Delta_{x} (\sin k_x+ i\sin k_y)
$ for the triplet case (Case II).

In Fig. 1, the relative phase of the Bloch function,
${\rm Arg}\, \frac {|\mb{k}\rangle_1 }{|\mb{k}\rangle_2}$,  for the singlet
case is shown.
For small $\Delta_0$,
one can see the charge two vortices at the pole of 
$\frac {|\mb{k}_1 \rangle}{|\mb{k}_2 \rangle} $ (circle) split into 
two independent ones and contribute to the non  trivial Chern
numbers.
Note that only the poles ( circles) of 
$
 \frac {|\mb{k}\rangle_1 }{|\mb{k}\rangle_2 }
$ contribute to the Chern numbers although the 
zeros (squares) also give vortices.
When $\Delta_0$ is sufficiently large,
they finally annihilate in pairs with negative vortices defined
by the zeros of $\frac {|\mb{k}_1 \rangle}{|\mb{k}_2 \rangle} $ (squares).
Another topological objects, the closed surface $R(T^2)$ and 
the Dirac string $z_-$ is shown for the singlet case in 
Fig.2. 
One can see the covering degree of mapping $T^2\to R(T^2)$ is two
if $\Delta_0$ is sufficiently small. 
The intersection number of the closed surface $R(T^2)$ and
the Dirac string $z_-$ is also two which can be easily checked 
for this example. 
In general, the order parameter may have more complicated momentum
dependence,
and  calculating the covering degree of mapping is not 
trivial at all. 
Therefore the intersection number 
$ I({R(T^2)},z_-)$ 
helps us to determine the Chern number concretely.
It is also easy to see that the Chern number vanishes
if $|\Delta_0|$ or $|\mu|$ is sufficiently large
since the surface $R(T^2)$ does not include the origin. 
In this singlet case, 
the origin $O$, is doubly covered since 
$ \mb{R} (\mb{k} )= \mb{R} (-\mb{k} )$
( $\Delta(\mb{k})= \Delta(-\mb{k}) $ ).
It implies the selection rule $\Delta C = \pm 2$\cite{ym-yh-pairing}.
Another interesting case is given by the $\Delta_{xy}=0$. 
In this case, the energy gap collapses and the gapless
Dirac dispersion prevent from determining the Chern number 
without ambiguity (if $|\mu|$ is small). 
This situation is evident by the present discussion since
the surface $R(T^2)$ is collapsed into a diamond shaped two dimensional
region 
$
\mb{R}  = 
\lambda_1 \mb{v}_1 +
\lambda_2 \mb{v}_2 -(0,0,\mu)$, $\lambda_{1,2}\in [-1,1]$,
$\mb{v} _1 = (\Delta_{x^2-y^2},0,-2t) $ 
and 
$\mb{v} _2 = (-\Delta_{x^2-y^2},0,2t) $. 
When the origin $O$ is on this region, 
the Chern number is ill-defined and the dispersion
is gapless. This condition is clearly given by 
$|\mu|\le 2t$. 
Otherwise, there is an energy gap and the Chern number
is zero.

The triplet case, the case II, gives a little tricky situation.
As shown in Fig. 3, the covering degree of mapping around the origin
is $+1$ or $-1$ depending on the parameter detail. 
The surface $R(T^2)$ is self-intersecting and a local 
 coordinate 
$
( \mb{e}_{x}( \mb{R}) ,\mb{e}_y (\mb{R})  , \mb{n} (\mb{R} ))
$
on the surface has a different chirality depending on the position of 
the surface
where 
$
\mb{e}_{\alpha}( \mb{R})=\widehat{
{\frac {\partial \mb{R}}{\partial k_{\alpha} } }}
$, $(\alpha=x,y)$ are tangent vectors
 and $\mb{n}(\mb{R} ) $ is a normal vector 
of the surface at $\mb{R} $.
In other words, inside and outside of the surface are exchanged 
on intersecting lines.
 Of course, the surface $R(T^2)$ itself is orientable 
always. 
At the transition, the selection rule for the 
triplet case is $\Delta C=\pm1$ generically.
However, the transition in which the origin $O$ passes through this
intersecting point, say, changing the chemical potential $\mu$, 
the Chern number can  change its sign as 
$C:\pm 1\to \mp 1$, $ (\Delta C = \pm 2) $.

In conclusion, we have demonstrated the topological quantum phase
transition in superconductivity on lattices by using several topological 
expressions for the Chern number.
They are the winding number of the 
relative phase of the quasiparticle Bloch function
$N_{winding}$ and 
the covering degree $N_{covering}$ 
of mapping $T^2\to R(T^2)$ around the Dirac monopole at origin.
The winding number also has another expression given by
the intersection number $I(R(T^2),z_-)$
of the closed surface $R(T^2)$ and the Dirac string $z_-$.

We thank Y. Morita and Y. Seki for discussions.

\vfill


%

\begin{figure}
\begin{center}
\includegraphics[width=6cm,clip]{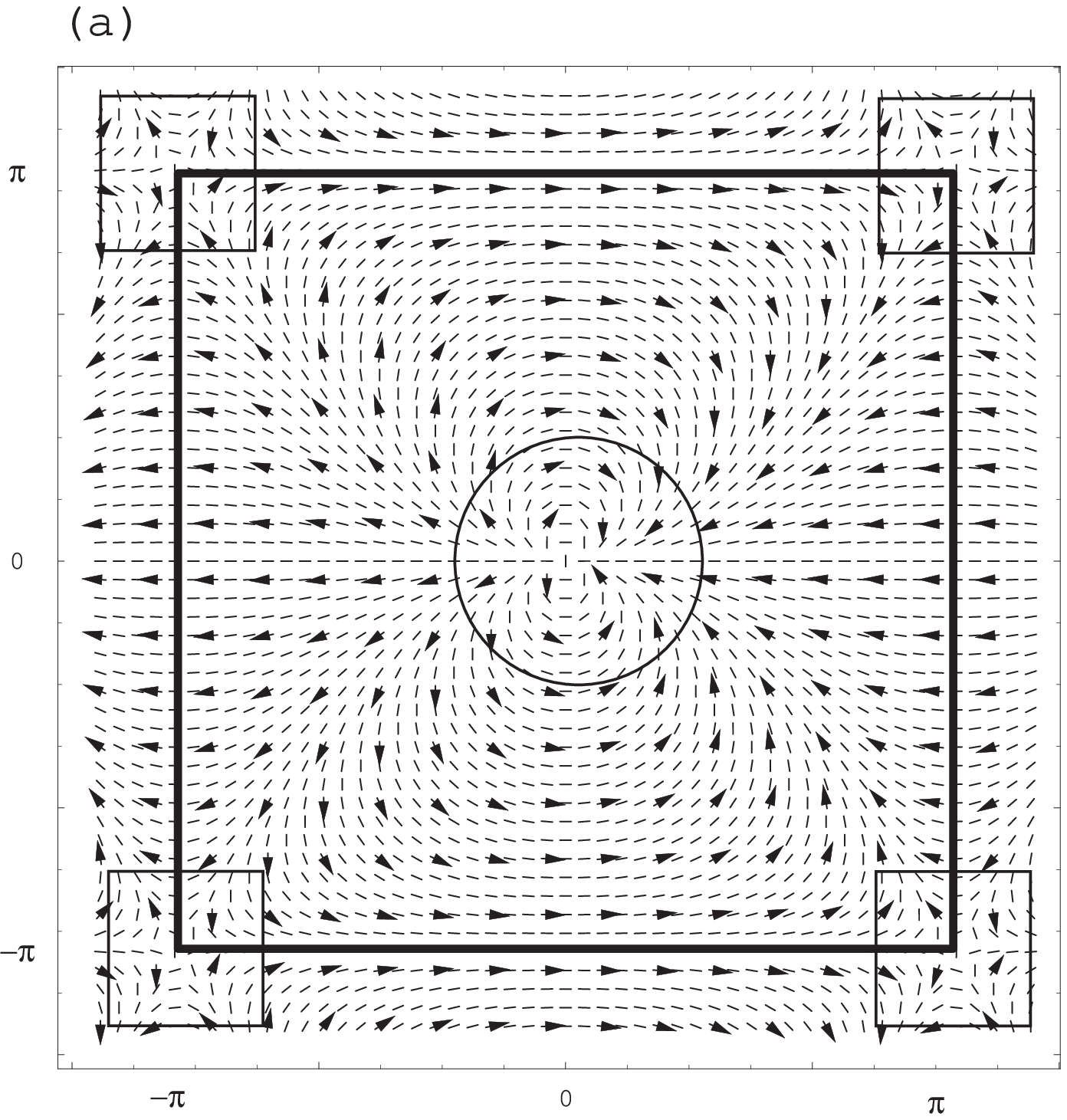}%
\end{center}
\begin{center}
\includegraphics[width=6cm,clip]{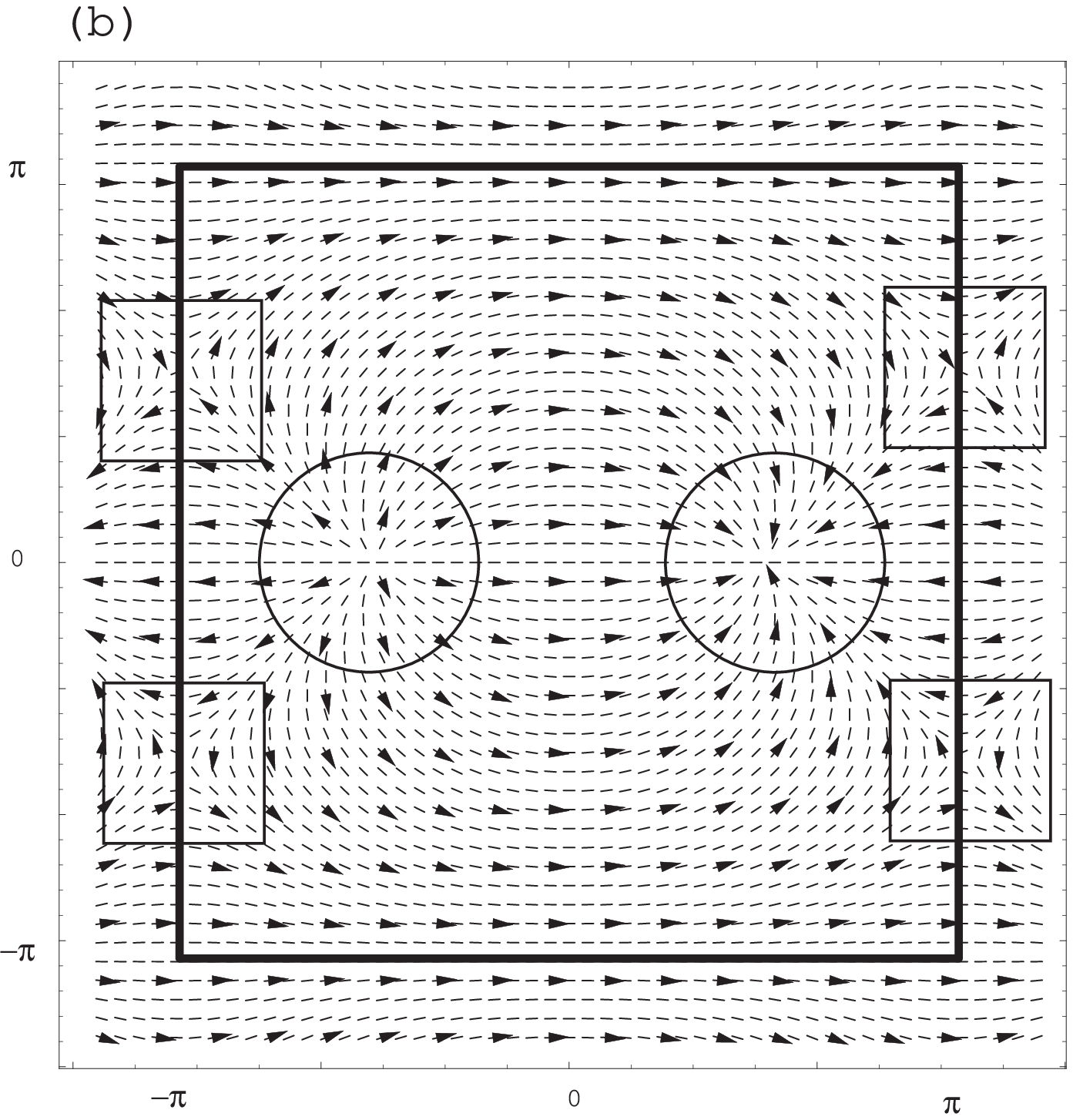}%
\end{center}
 \caption{Relative phases, 
${\rm Arg}\,  \frac {| \mb{k} \rangle_1}{ {| \mb{k} \rangle_2}}
={\rm Im}\,\log  \frac {| \mb{k} \rangle_1}{ {| \mb{k} \rangle_2}} $ 
of the quasiparticle Bloch function
in the Brillouin zone ( solid square ). 
The circles and the squares include the poles 
and the zeros of 
of 
$ \frac {|\mb{k}  \rangle_1}{
 {|\mb{k} \rangle_2}}$
 respectively.
Note that only the circles ( poles ) contribute to the 
total Chern number.
The spin singlet case
with (a)
$\Delta_0=0,\ \Delta_{x^2-y^2}=t,\ \Delta_{xy}=0.5t,\ \mu=0$
(the vorticity of the pole at the origin is $+2$),
(b)
$\Delta_0=2t,\ \Delta_{x^2-y^2}=t,\ \Delta_{xy}=0.5t,\ \mu=0$
( there are two vorticity $+1$ vortices at the poles).
\label{fig1ab}}
 \end{figure}

\begin{widetext}

\begin{figure}
\begin{center}
\end{center}
 \caption{Mapped Brillouin zone $R(T^2)$ by the 
Berry's parametrization.
The monopole at the origin $O$ and the Dirac string $z_-$ are
also shown.
The spin singlet case
with
$\Delta  _0=0,\ \Delta_{x^2-y^2}=t,\ \Delta_{xy}=t,\ \mu=-t$.
(a) is the total surface $R(T^2)$, $k_x\in {[}0,2\pi{]}$, $k_y\in {[}0,2\pi{]}$
and (b)-(f) are parts of the surface to show how it covers the origin.
They are drawn for $k_x\in {[}0,2\pi{]}$ and
(b): $k_y\in {[} 0,\frac {2\pi}{5}{]}$,  
(c): $k_y\in {[} \frac {2\pi}{5},\frac {4\pi}{5}{]}$,  
(d): $k_y\in {[} \frac {4\pi}{5},\frac {6\pi}{5}{]}$,  
(e): $k_y\in {[} \frac {6\pi}{5},\frac {8\pi}{5}{]}$,  
(f): $k_y\in {[} \frac {8\pi}{5},2\pi{]}$.
The monopole $O$ is doubly covered by surface $R(T^2)$.
\label{fig2abc}}
\end{figure}

\begin{figure}
\begin{center}
\end{center}
 \caption{
Same as Fig.2 for the spin triplet  case
with
$\Delta  _0=0,\ \Delta_{x}=-t,\ \mu=-t$.
(a) is the total surface $R(T^2)$, $k_x\in {[}0,2\pi{]}$, $k_y\in {[}0,2\pi{]}$.
(b) and (c) are drawn for $k_x\in {[}0,2\pi{]}$ and
(b): $k_y\in {[}0,\pi{]}$, (c): $k_y\in {[}\pi,2\pi{]}$.
\label{fig3abc}}
 \end{figure}


\end{widetext}

%






\end{document}